\documentclass[runningheads]{llncs}

\usepackage[T1]{fontenc}
\usepackage{amsmath}
\usepackage{graphicx}
\usepackage{booktabs}
\usepackage{array}
\usepackage{tabularx}
\usepackage{multirow}
\usepackage{xcolor}
\PassOptionsToPackage{hyphens}{url}
\usepackage{hyperref}
\usepackage{xurl}
\usepackage{tikz}
\usetikzlibrary{positioning,arrows.meta,shapes.geometric,fit,backgrounds}
\usepackage{enumitem}
\setlist[itemize]{leftmargin=*,topsep=2pt,itemsep=1pt}
\setlist[enumerate]{leftmargin=*,topsep=2pt,itemsep=1pt}

\hypersetup{
  colorlinks=true,
  linkcolor=black,
  citecolor=black,
  urlcolor=black
}

\begin{document}

\title{Regulating the Machine Contributor:\\Governance and Policy Alignment in Open Source}

\titlerunning{Regulating the Machine Contributor}

\author{Jassem Manita\inst{1}\orcidID{0009-0009-1558-3855} \and
        Aziz Amari\inst{2}\orcidID{0009-0005-7020-3051}}

\authorrunning{J. Manita and A. Amari}

\institute{
Faculty of Sciences of Tunis (FST), University of Tunis El Manar, Tunis, Tunisia\\
\email{jassem.manita@etudiant-fst.utm.tn}
\and
National Institute of Applied Science and Technology (INSAT),\\University of Carthage, Tunis, Tunisia\\
\email{aziz.amari@insat.ucar.tn}}

\maketitle

\begin{abstract}
AI-assisted software development has moved from line-level autocomplete to agents that can plan changes, edit files, and submit pull requests with limited human supervision. Open-source software, however, evolves through a process designed for humans: contributor agreements, codes of conduct, and review norms all assume a legally accountable person who can attest to provenance and answer reviewer questions. Autonomous and semi-autonomous AI contributors strain those assumptions, and the 2025--2026 record of agent-driven incidents, AI-generated nuisance volume, and platform-level shutdowns shows that the gap is operationally consequential. Several open-source organisations have responded with contribution policies, but the result is fragmented, and its alignment with emerging AI governance frameworks (EU AI Act, NIST AI RMF with the UC Berkeley Agentic AI Profile, ISO/IEC 42001 and 23894) is unmapped at the contribution level. We compare policies across six organisations (SymPy, LLVM, matplotlib, OpenInfra, the Apache Software Foundation, and the Linux Foundation) using Most-Similar Systems Design with indicator-based coding and process tracing for SymPy and LLVM. From this we derive a six-dimensional taxonomy (disclosure, responsibility, human oversight, licensing, enforcement, maintainer workload), an ordinal Policy Maturity Score, and a mapping of documented agent incidents onto the dimensions each policy fails to govern. Aligning the dimensions with the regulatory frameworks above identifies overlapping gaps neither side currently closes, and we close by sketching the shape of a harmonised tiered framework and the empirical evaluation needed to calibrate it.

\end{abstract}

%======================================================================
\section{Introduction}\label{sec:intro}
%======================================================================

AI-assisted software development has changed quickly. Coding tools that started as line-level completion engines now operate as agents that can plan changes, edit multiple files, run tests, and submit pull requests with limited human supervision. Open-source software, which underpins much of modern infrastructure, evolves through a process designed for a different actor: a person proposes a change, another person reviews it, and human judgement carries both sides of the exchange. Open-source projects encode this assumption in their governance instruments. Contributor License Agreements (CLAs), Developer Certificates of Origin (DCOs), codes of conduct, and review norms all assume a legally accountable human contributor who can attest to provenance, answer reviewer questions, and bear responsibility for downstream effects. None of these instruments fit an AI agent acting without per-contribution human approval: agents have no legal standing to make warranties, no insurable liability, and no mechanism for downstream remediation if their contributions cause harm. The mismatch became operationally consequential in 2025--2026.

The visible failures span several modes. In February 2026 an autonomous agent operating on the OpenClaw platform under the handle \texttt{crabby-rathbun} submitted pull requests to matplotlib and SymPy and, after matplotlib closed the first under its standing prohibition on autonomous-agent contributions, published a blog post attacking the maintainer by name \cite{shambaugh2026,sympy_pr29156}. SymPy's October 2025 mailing-list thread documented rising AI-generated PR volume \cite{sympy_mailinglist}; LLVM cited the same pattern as the rationale for its new AI-tool policy \cite{llvm_policy}; curl shut down its HackerOne bug bounty in February 2026 over low-quality AI-generated reports \cite{stenberg_curl_hackerone}; and SecurityScorecard's STRIKE team identified more than 41{,}000 exposed instances of the OpenClaw agent platform \cite{strike_openclaw}. The common signal across these events is that AI lowers the cost of producing contributions but does not lower the cost of trustworthy human review.

Several open-source organisations have written contribution policies that try to address this gap. SymPy \cite{sympy_policy}, LLVM \cite{llvm_policy}, matplotlib \cite{matplotlib_policy}, OpenInfra \cite{openinfra_policy}, the Apache Software Foundation \cite{apache_policy}, and the Linux Foundation \cite{lf_policy} each publish guidance on AI-assisted or AI-generated contributions, but the result is fragmented: some prohibit autonomous-agent contributions outright, others only require disclosure, others impose human-in-the-loop standards, others address licensing provenance alone. Meanwhile, formal AI governance frameworks have begun to bind providers and deployers of AI systems: the EU AI Act \cite{euaiact}, the NIST AI Risk Management Framework with the UC Berkeley Agentic AI Profile \cite{nistaimrmf,ucberkeley_agentic}, and ISO/IEC 42001 \cite{iso42001} and 23894 \cite{iso23894}. Open-source contribution is where those frameworks meet operational reality, so mapping community policies against them shows where community practice already exceeds regulation, where regulation exposes gaps the policies leave open, and where neither side offers guidance.

This paper presents that comparison and mapping, organised around three research questions:
\begin{description}
\item[\textbf{RQ1}] What dimensions do existing AI-contribution policies address, and what spectrum of approaches exists within each dimension?
\item[\textbf{RQ2}] Where do policies converge toward common approaches, and where do they diverge in ways that reflect genuine governance choices?
\item[\textbf{RQ3}] How do open-source AI-contribution policies map onto requirements in the EU AI Act, NIST AI RMF, and ISO standards, and what gaps exist in either direction?
\end{description}
We use Most-Similar Systems Design with indicator-based coding across the six cases and process tracing for SymPy and LLVM, and make five contributions: (i) four AI contribution modes that keep distinct governance questions apart; (ii) a six-dimensional taxonomy (disclosure, responsibility, human oversight, licensing, enforcement, maintainer workload); (iii) an ordinal Policy Maturity Score that locates each case on the spectrum; (iv) two policy archetypes, licensing-first and oversight-first, that solve different problems rather than marking weak and strong points on one scale; and (v) a regulatory alignment that identifies overlapping gaps and shows maintainer workload to be the dimension neither policies nor frameworks address. We close by sketching a tiered framework and the evaluation needed to calibrate it, stopping short of a calibrated v1 the evidence does not yet support.

%======================================================================
\section{Background and Motivation}\label{sec:background}
%======================================================================

\subsection{AI Contribution Modes}\label{sec:modes}

Policy-relevant questions differ across the autonomy spectrum, and collapsing AI involvement into one category is the most common source of policy error. We distinguish four modes used throughout the paper: \emph{AI-assisted human contribution}, where a human uses an AI tool as an authoring aid and remains responsible; \emph{AI-generated contribution}, describing substantive content produced by an AI system but still submitted and reviewed by a human (provenance, not autonomy); \emph{semi-autonomous agent contribution}, an agentic workflow where an AI performs multi-step tasks but a human gates the final submission; and \emph{autonomous agent contribution}, where an AI agent opens issues, PRs, or comments without meaningful per-action human approval. The crabby-rathbun/OpenClaw matplotlib and SymPy incidents are the central open-source examples of the fourth mode. A policy that addresses only one of these modes leaves the others uncovered.

\subsection{Documented Agent Incidents (2025--2026)}

The window from October 2025 through February 2026 produced the first wave of documented harm involving autonomous AI contributors in open-source ecosystems. We organise the record by failure pattern.

\textbf{Pattern 1: no human checkpoint before contribution actions.} An OpenClaw-hosted agent operating as \texttt{crabby-rathbun} opened matplotlib PR \#31132 on 9 February 2026 and SymPy PR \#29145 two days later, neither preceded by a human attesting to scope, quality, or licensing position \cite{shambaugh2026,sympy_pr29156}. matplotlib maintainer Scott Shambaugh closed PR \#31132 on 11 February 2026 under matplotlib's standing prohibition on autonomous-agent contributions; SymPy contributors recognised the same account on the SymPy PR a day later and flagged it on the project mailing list.

\textbf{Pattern 2: scale outpacing review capacity.} The same window shows volume outrunning review: the SymPy thread of 26 October 2025 \cite{sympy_mailinglist}, LLVM citing rising nuisance volume as its policy rationale \cite{llvm_policy}, curl's HackerOne shutdown on 1 February 2026 after low-quality AI reports made triage unsustainable \cite{stenberg_curl_hackerone}, and SecurityScorecard STRIKE's count of more than 41{,}000 exposed OpenClaw instances \cite{strike_openclaw}. Enough infrastructure is deployed that the gap between agent capability and review capacity is widening.

\textbf{Pattern 3: agent-generated harm to people who are not users of the system.} After matplotlib closed PR \#31132, crabby-rathbun produced and distributed a blog post titled \emph{``Gatekeeping in Open Source: The Scott Shambaugh Story''}, accusing the maintainer of insecurity and of ``protecting his little fiefdom'' \cite{shambaugh2026}. The post was circulated across GitHub and the agent's own site.

The SymPy community diagnosed the underlying coverage gap directly on its mailing list following the incident: ``Our current AI policy covers humans using AI tools assuming that a human is operating and is responsible for the contribution. In the agent case no human will be held accountable for what it's doing'' \cite{sympy_pr29156}. The crabby-rathbun case instantiates all three patterns simultaneously; Pattern 2 confirms that volume-driven review burden is an ecosystem issue rather than a single event.

\subsection{Regulatory Frameworks}

Three regulatory instruments are directly relevant to AI-contribution governance.

The \textbf{EU AI Act} \cite{euaiact} is the only legally binding framework in this set. Article 13 requires transparency sufficient for deployers to interpret outputs; Article 14 requires that natural persons be able to oversee AI systems, interpret their outputs, and intervene; Articles 16--29 establish provider/deployer accountability distinctions; Article 5(1)(b) prohibits AI systems that ``exploit any of the vulnerabilities of a natural person'' to influence behaviour. Industry analysis notes a known ambiguity: the Act does not define what separates meaningful oversight from rubber-stamping \cite{iapp_oversight}.

The \textbf{NIST AI Risk Management Framework} \cite{nistaimrmf} is voluntary in the United States but functions as the dominant operational AI-governance instrument internationally. The UC Berkeley CLTC \emph{Agentic AI Risk-Management Standards Profile} of February 2026 \cite{ucberkeley_agentic} extends the NIST four-function structure (\textsc{Govern}, \textsc{Map}, \textsc{Measure}, \textsc{Manage}) specifically to autonomous agents. The Berkeley profile names \emph{anthropomorphic and socially persuasive behaviour} as a measurable risk; this gives the class of behaviour exhibited by crabby-rathbun a standard label it would otherwise lack.

\textbf{ISO/IEC 42001} (AI management systems) \cite{iso42001} and \textbf{ISO/IEC 23894} (AI risk management) \cite{iso23894} are not certification requirements for open-source projects, but they set the conditions under which a contribution policy is operational rather than aspirational: ISO 42001 separates a policy statement from a management system with documented roles, and ISO 23894 requires risk assessment at every lifecycle stage. Both expose structural gaps that policy text alone cannot close.

%======================================================================
\section{Methodology}\label{sec:methods}
%======================================================================

Our analysis combines Comparative Policy Analysis as the frame (\S\ref{sec:cpa}), Most-Similar Systems Design for case selection (\S\ref{sec:mssd}), indicator-based coding of policy text on six dimensions (\S\ref{sec:coding}), process tracing for two cases (\S\ref{sec:proctrace}), and an ordinal Policy Maturity Score (\S\ref{sec:scoring}). Results follow in Section~\ref{sec:taxonomy}.

\subsection{Comparative Policy Analysis as the Right Frame}\label{sec:cpa}

We compare policies against a common set of dimensions and build a taxonomy that supports cross-case reasoning and regulatory alignment, the standard task of Comparative Policy Analysis (CPA) \cite{peters2020handbook,geva2018twenty}.

The hardest methodological challenge is what the literature calls the \emph{common-variable problem} \cite{wallis2021conceptual}: how to identify genuinely comparable dimensions when policies use different language, structures, or institutional forms. SymPy requires contributors to ``understand'' AI-generated code; LLVM requires a human to ``take responsibility''; matplotlib requires contributors to ``fully understand the proposed changes and \dots\ explain why they are the correct approach''; Apache focuses on licensing provenance with no understanding requirement at all. All four address the same governance question through structurally different mechanisms. CPA best practice is to define analytic dimensions \emph{a priori}, grounded in theory, existing literature, and regulatory frameworks, and use them as a common coding lens \cite{pasetti2024indicator}. We define our six dimensions before coding, not as a byproduct of reading the policies.

\subsection{Most-Similar Systems Design (MSSD)}\label{sec:mssd}

Most-Similar Systems Design \cite{anckar2020mssd} selects cases similar in context but differing on the variable of interest, here the policy itself, so differences trace to governance choices rather than resources or capacity. Our six cases qualify: all are major open-source projects or foundations, all set their current positions during the 2023--2025 rise in AI-generated contributions, and all faced the same triggering problem.

Each case anchors a distinct position (Table~\ref{tab:cases}). Two further organisations are retained as \emph{validation cases}: CPython/PSF, which has no AI-specific policy as of February 2026 and serves as a policy-absence reference, and SAP, whose corporate governance would confound policy differences with structural ones if used as a primary case.

\begin{table}[htbp]
\centering
\caption{Primary cases and the analytic position each anchors.}\label{tab:cases}
\small
\renewcommand{\arraystretch}{1.2}
\begin{tabularx}{\textwidth}{@{}>{\bfseries}lX@{}}
\toprule
Case & \textbf{Anchored position} \\
\midrule
SymPy & Anticipatory community governance; policy drafted through mailing-list mobilisation \emph{before} adversarial incident \cite{sympy_policy} \\
\addlinespace[3pt]
LLVM & Maximum human-oversight requirement; contributor must answer questions during review without referring to the AI; ``Good First Issue'' tickets excluded \cite{llvm_policy} \\
\addlinespace[3pt]
matplotlib & Strongest enforcement language: explicit autonomous-agent prohibition with ban + GitHub report \cite{matplotlib_policy} \\
\addlinespace[3pt]
OpenInfra & Most comprehensive policy: structured labelling (\texttt{Generated-By:}/\texttt{Assisted-By:}), human-in-loop, understanding requirement, reviewer scrutiny, licensing guidance \cite{openinfra_policy} \\
\addlinespace[3pt]
Apache (ASF) & Licensing-first; originator of the \texttt{Generated-By:} label adopted by OpenInfra and others \cite{apache_policy} \\
\addlinespace[3pt]
Linux Foundation & Ecosystem coordinator; licensing guidance focused on GPL contamination, tool ToU compliance, copyright provenance \cite{lf_policy} \\
\bottomrule
\end{tabularx}
\end{table}

\subsection{Indicator-Based Coding: Six Dimensions}\label{sec:coding}

The six dimensions are defined \emph{a priori} from regulatory frameworks rather than derived inductively from cases:

\begin{description}[leftmargin=2em,style=nextline]
\item[D1 \textsc{Disclosure}] Whether and how contributors must declare AI tool use, which specific tools, and to what degree outputs were modified. Maps to EU AI Act Article 13 \cite{euaiact}.
\item[D2 \textsc{Responsibility}] Who is accountable for AI-generated content, and whether autonomous agents are addressed separately from humans-using-AI. Maps to EU AI Act Articles 16--29 (provider/deployer) \cite{euaiact} and the Berkeley Agentic Profile \textsc{Govern} function \cite{ucberkeley_agentic}.
\item[D3 \textsc{Human Oversight}] Required review standard, explainability obligation, and answerability of the human contributor. Maps to EU AI Act Article 14 \cite{euaiact} and Berkeley \textsc{Map} (autonomy-proportional control).
\item[D4 \textsc{Licensing}] Treatment of copyright, terms-of-use compatibility, training-data provenance, and GPL contamination risk. Maps to EU AI Act Article 53 (GPAI transparency) at the model level; underspecified at the contributor level.
\item[D5 \textsc{Enforcement}] Compliance verification mechanisms and consequences. Maps to ISO 42001 management-system requirements \cite{iso42001}.
\item[D6 \textsc{Maintainer Workload}] Whether reviewer cognitive burden is recognised and structurally addressed. \emph{No regulatory framework currently addresses this dimension at all}; this is the most original element of the taxonomy.
\end{description}

For each (case, dimension) pair we code (i) the assigned category, (ii) direct supporting language from the policy text, and (iii) the source URL. Where a dimension is unaddressed, the cell is coded \textsc{Absent} explicitly, with the absence categorised as either \emph{structural} (deliberately out of scope) or \emph{implicit} (not mentioned without rationale).

\subsection{Process Tracing for Two Cases}\label{sec:proctrace}

Indicator coding shows what differs across the policies, not why each took its form. Process tracing \cite{blatter2012designing} reconstructs that causal chain (\S\ref{sec:tracing}); we apply it to SymPy and LLVM, whose formation records are public and detailed. matplotlib's and OpenInfra's are not traceable, and Apache's and the Linux Foundation's documents already state their rationale.

\subsection{Ordinal Scoring}\label{sec:scoring}

To enable cross-case comparison and to map incidents back onto policy gaps (Section~\ref{sec:incidents}), we layer an ordinal score $s \in \{0,1,2,3,4,5\}$ on top of each coded cell using the rubric in Table~\ref{tab:rubric}. Aggregate scores are reported as a \emph{Policy Maturity Score} (PMS, sum of dimension scores, $\text{max}=30$).

\begin{table}[htbp]
\centering
\caption{Ordinal coding rubric, applied per (case, dimension) cell.}\label{tab:rubric}
\small
\begin{tabularx}{\textwidth}{@{}cX@{}}
\toprule
\textbf{Score} & \textbf{Interpretation} \\
\midrule
0 & \textsc{Absent}: dimension not addressed by the policy text at all. \\
1 & \textsc{Acknowledged}: dimension referenced normatively (rationale only); no operational provision. \\
2 & \textsc{Recommended}: optional or aspirational provision with no enforcement mechanism. \\
3 & \textsc{Mandatory (general)}: provision is mandatory but does not distinguish autonomous agents from humans-using-AI. \\
4 & \textsc{Mandatory (agent-aware)}: provision is mandatory \emph{and} explicitly addresses autonomous-agent scenarios. \\
5 & \textsc{Operational + verifiable}: provision is mandatory, agent-aware, and includes a verification or enforcement mechanism. \\
\bottomrule
\end{tabularx}
\end{table}

%======================================================================
\section{Findings}\label{sec:taxonomy}
%======================================================================

Table~\ref{tab:grid} shows the coded grid and the resulting Policy Maturity Scores; each subsection then gives one dimension's key finding and its regulatory mapping.

\begin{table}[htbp]
\centering
\caption{Indicator-coded grid. Each cell shows the coded category and ordinal score $s\in\{0,\dots,5\}$. PMS is the column sum (max 30).}\label{tab:grid}
\scriptsize
\setlength{\tabcolsep}{3pt}
\resizebox{\textwidth}{!}{%
\begin{tabular}{@{}c l l l l l l@{}}
\toprule
 & \textbf{SymPy} & \textbf{LLVM} & \textbf{matplotlib} & \textbf{OpenInfra} & \textbf{Apache} & \textbf{Linux Fdn.} \\
\midrule
D1 & Mand.\ (3)        & Mand.+lbl (4)     & Cond.\ (2)        & Struct.\ (5)      & Recom.\ (2)       & Absent (0) \\
D2 & agt gap (2)       & agt ban (4)       & agt ban (4)       & no ban (3)        & CLA (2)           & frame (2) \\
D3 & Und+def (3)       & Und+ans (5)       & Und+v-add (4)     & Und+scr (4)       & Absent (0)        & Absent (0) \\
D4 & Abs.\ (AI) (1)    & Gen.\ cop. (2)    & Abs.\ (AI) (1)    & Compr.\ (4)       & Compr.\ (4)       & Compr.\ (4) \\
D5 & Com.\ (2)         & M.-led (3)        & b+r (5)           & Com.\ (2)         & Com.\ (2)         & S. del.\ (1) \\
D6 & Part.\ nor.\ (1)  & Part.\ rat.\ (2)  & Part.\ exp.\ (2)  & Absent (0)        & Absent (0)        & Absent (0) \\
\midrule
\textbf{PMS} & \textbf{12} & \textbf{20} & \textbf{18} & \textbf{18} & \textbf{10} & \textbf{7} \\
\bottomrule
\end{tabular}%
}

\vspace{2pt}
{\scriptsize \emph{Key:} agt = autonomous agent; und = understanding; ans = answerability without AI; v-add = value-add; scr = reviewer scrutiny; b+r = ban + report; com. = community norm; m.-led = maintainer-led; s.\ del. = structurally delegated; abs.\ (AI) = no AI-specific provision; nor./rat./exp. = normative / rationale-only / explicit.}
\end{table}

For each dimension we report the key finding and its regulatory mapping below.

\subsection{D1 Disclosure: The Disclosure Reversal}\label{sec:d1}
Apache originated the \texttt{Generated-By:} label in June 2023 as an explicitly voluntary recommendation \cite{apache_policy}. OpenInfra adopted the same lexical instrument and made it mandatory, adding a second tier (\texttt{Assisted-By:}) to capture the assistive end of the autonomy spectrum \cite{openinfra_policy}. The adopter is stricter than the originator, a policy-diffusion pattern that runs counter to the typical assumption that diffusion dilutes commitments. Apache's 2023 document explicitly acknowledged its own limitations and the need for future revision; OpenInfra's adoption is exactly that revision, executed externally rather than internally. Apache's original guidance frames the practice as optional:
\begin{quote}
When providing contributions authored using generative AI tooling, a recommended practice is for contributors to indicate the tooling used to create the contribution. This should be included as a token in the source control commit message, for example including the phrase ``Generated-by:''. This allows for future release tooling to be considered that pulls this content into a machine parsable Tooling-Provenance file.
\end{quote}\cite{apache_policy}
OpenInfra's 2025 revision makes the same label mandatory and adds verification during review, making disclosure checkable at the contribution level \cite{openinfra_policy}. \emph{Regulatory mapping}: EU AI Act Article 13 (transparency for high-risk systems) is the upstream anchor; Apache's and OpenInfra's labelling instruments operationalise Article 13 at the contribution level more concretely than the Act addresses transparency at the system level.

\subsection{D2 Responsibility: The Autonomous-Agent Gap}
Only LLVM and matplotlib explicitly address the autonomous-agent scenario. The other four assign responsibility to ``the contributor'' without addressing what happens when no human contributor exists. The crabby-rathbun incident exposed this gap operationally: matplotlib's policy applied (the agent's PR was rejected under the existing prohibition), but SymPy's policy did not, prompting Manita's Issue \#29155 \cite{sympy_issue29155} and proposed PR~\#29156 amendment on 13 February 2026 \cite{sympy_pr29156}; as of 7 May 2026, the issue remains open and the PR was closed unmerged. \emph{Regulatory mapping}: the EU AI Act provider/deployer distinction (Articles 16--29) defines accountability for AI systems but cannot resolve the case where the ``deployer'' is itself an autonomous agent without legal standing. The Berkeley Agentic Profile \textsc{Govern} function requires policies to specify the scope of permitted autonomous action \emph{before} it occurs \cite{ucberkeley_agentic}; only LLVM and matplotlib meet this requirement.

\subsection{D3 Human Oversight: The LLVM/Article-14 Paradox}
LLVM's policy states:
\begin{quote}
Contributors must read and review all LLM-generated code or text before they ask other project members to review it. The contributor is always the author and is fully accountable for their contributions. Contributors should be sufficiently confident that the contribution is high enough quality that asking for a review is a good use of scarce maintainer time, and they should be able to answer questions about their work during review.
\end{quote}\cite{llvm_policy}
This answerability requirement (the contributor must answer questions during review without referring back to the AI) is operationally more demanding than EU AI Act Article 14, which mandates oversight \emph{capacity} but not demonstrable understanding \cite{euaiact,iapp_oversight}. A community-governed open-source project has imposed a stricter oversight standard than the binding regulatory text governing high-risk AI systems in the European Union.

\subsection{D4 Licensing: The Licensing/Oversight Inverse}\label{sec:d4}
Apache and Linux Foundation: comprehensive licensing guidance, zero human-oversight requirements. SymPy and matplotlib: strong oversight, zero AI-specific licensing. LLVM and OpenInfra: partial coverage of both. SymPy's contributing guidance shows the oversight-first stance treating licensing as orthogonal to AI provenance:
\begin{quote}
All code in SymPy is released under the BSD 3-clause copyright license. Contributors to SymPy license their code under the same license when it is included into SymPy's version control repository. That means contributors must own the copyright of any code submitted to SymPy or must include the BSD 3-clause compatible open source license(s) associated with the submitted code in the patch.
\end{quote}\cite{sympy_policy}
The scatter is not a single spectrum but two archetypes: \emph{licensing-first foundation policy} (Apache, LF, driven by legal-liability concerns) and \emph{oversight-first community policy} (SymPy, matplotlib, driven by maintainer review burden), with LLVM and OpenInfra hybridising the two.

\subsection{D5 Enforcement}
Only matplotlib includes an explicit prohibition on agent accounts with consequence language. The contributing guide states:
\begin{quote}
Unacceptable uses: External AI tooling (e.g. bots, agents) directly interacting with the project; including creating issues, PRs or commenting on GitHub or Discourse.
\end{quote}\cite{matplotlib_policy}
and sets out enforcement expectations:
\begin{quote}
To ensure project health and preserve limited core developer capacity, we will flag and reject low-value contributions that we believe are AI generated. We may ban and/or report users to GitHub if they harm the project or its community through irresponsible use of AI.
\end{quote}\cite{matplotlib_policy}
It was, accordingly, the only policy operationally enforced against the crabby-rathbun agent (PR \#31132 was closed under it on 11 February 2026 \cite{shambaugh2026}). The remaining policies rely on reviewer discretion or community norms, which the same agent's repeated submissions demonstrate to be insufficient under sustained adversarial pressure.

\subsection{D6 Maintainer Workload: The Universal Gap}\label{sec:d6}
No policy in the set, and no regulatory framework reviewed (EU AI Act, NIST AI RMF, Berkeley Agentic Profile, ISO 42001, ISO 23894), provides a structural mechanism for protecting reviewer capacity against asymmetric submission volume. Three policies acknowledge the issue normatively; none address it mechanistically (no rate limits, cooldown periods, automated triage, or volume caps). The empirical record outside our case set confirms the cost: the SymPy mailing-list thread opened by Oscar Benjamin on 26 October 2025 \cite{sympy_mailinglist}; Daniel Stenberg's shutdown of curl's HackerOne bug-bounty programme on 1 February 2026, citing an unsustainable AI-generated submission volume (roughly 8$\times$ the normal rate at a 0\% verification rate, against a 15\%+ historical baseline) \cite{stenberg_curl_hackerone}; and the increase in LLM-assisted nuisance contributions referenced in LLVM's policy text itself \cite{llvm_policy}.

%======================================================================
\section{Discussion}\label{sec:discussion}
%======================================================================

The coded grid says where each policy sits but not how it got there, how it holds under adversarial pressure, or how it relates to the formal governance frameworks now binding providers and deployers. We take these in turn: process tracing for two policies (\S\ref{sec:tracing}), an incident-to-dimension mapping (\S\ref{sec:incidents}), and a regulatory-alignment analysis (\S\ref{sec:regulatory}).

\subsection{Process Tracing: How Two Policies Formed}\label{sec:tracing}

We trace SymPy and LLVM, the two cases with public formation records detailed enough to follow.

\subsubsection*{SymPy: anticipatory governance through community mobilisation.}

\textbf{Timeline.} Oscar Benjamin opened a mailing-list thread on 26 October 2025 documenting rising low-quality AI-generated PRs \cite{sympy_mailinglist}; Jason Moore drafted the first policy in PR \#28941 (14 January 2026), merged weeks before the incident \cite{sympy_pr28941}. When crabby-rathbun submitted PR \#29145 on 12 February, the community found the merged policy covered humans-using-AI but not autonomous agents, and Manita opened Issue \#29155 \cite{sympy_issue29155} and PR \#29156 the next day \cite{sympy_pr29156}. As of 7 May 2026 the issue is open and the PR closed unmerged; the gap is stress-tested and unresolved.

\textbf{Causal chain.} The SymPy policy emerged from sustained mailing-list mobilisation rather than from a single triggering incident. The community formalised a policy in anticipation of an adversarial case it had not yet encountered, and then encountered it. The policy held in spirit (the agent's PR was flagged and rejected), but the literal text needed revision because it was scoped to humans operating tools, not autonomous agents. The post-incident revision is therefore an instance of \emph{policy-by-stress-test}: an existing policy is exercised by an adversarial case, the gap is documented, and the next iteration is informed by primary-source evidence.

\textbf{Key decisions.} The understanding-and-defence requirement and the omission of AI-specific licensing were both deliberate (the project's BSD licence already governs every contribution); the missing autonomous-agent provision was not deliberate, and surfaced post-hoc as a coverage gap.

\subsubsection*{LLVM: institutional reasoning from observed patterns.}

\textbf{Context.} LLVM's policy text states the formation rationale directly: ``Over the course of 2025, we observed an increase in the volume of LLM-assisted nuisance contributions'' \cite{llvm_policy}. The triggering institutional event was a specific high-profile PR that received significant attention on Hacker News, which catalysed Reid Kleckner's policy proposal.

\textbf{Causal chain.} LLVM's pathway differs structurally from SymPy's: not a sustained mailing-list mobilisation but an episodic pattern that reached a tipping point through external visibility. The result is the most operationally demanding policy in the set: contributors must answer questions about AI-generated code during review without referring back to the AI; AI-assisted contributions must be explicitly labelled with an \texttt{Assisted-by:} trailer; autonomous agents are explicitly prohibited; and ``Good First Issue'' tickets cannot be used as AI tasks (a deliberate carve-out to preserve onboarding pathways for new human contributors).

\textbf{Key decisions.} All three choices were deliberate, and the autonomous-agent prohibition predated the crabby-rathbun incident, anticipating the scenario rather than reacting to it.

The two pathways differ: SymPy deliberated and then stress-tested its policy; LLVM reasoned from an observed pattern, anticipating rather than reacting. Both are credible, but neither is universally available: one needs mailing-list bandwidth, the other institutional infrastructure. Projects with neither lack a defensive default, which motivates the framework sketch in Section~\ref{sec:framework}.

\subsection{Documented Incidents Against the Dimensions}\label{sec:incidents}

The taxonomy is descriptive unless it can be put to work explaining incidents. Table~\ref{tab:incidents} maps the documented 2025--2026 open-source agent incidents to the dimensions whose absence each one exposes. The mapping is conservative: a dimension is marked only when a structural failure in the public account matches a coded gap in at least one case. We restrict the corpus to harms whose locus is the contribution channel itself; broader agent incidents (deepfake fraud, prompt-injection, agentic UI failures, robotaxi events) fall outside contribution-policy analysis.

\begin{table}[htbp]
\centering
\caption{Documented 2025--2026 open-source agent incidents mapped to policy dimensions. A bullet marks an incident whose failure mode corresponds to a coded gap on that dimension in at least one case.}\label{tab:incidents}
\footnotesize
\setlength{\tabcolsep}{6pt}
\begin{tabular}{@{}p{6.6cm}cccccc@{}}
\toprule
\textbf{Incident} & \textbf{D1} & \textbf{D2} & \textbf{D3} & \textbf{D4} & \textbf{D5} & \textbf{D6} \\
\midrule
crabby-rathbun matplotlib PR \#31132 \cite{shambaugh2026}             & $\bullet$ & $\bullet$ & $\bullet$ &           & $\bullet$ & $\bullet$ \\
crabby-rathbun SymPy PR \#29145 \cite{sympy_pr29156}                  & $\bullet$ & $\bullet$ & $\bullet$ &           & $\bullet$ & $\bullet$ \\
SymPy nuisance-PR mobilisation thread \cite{sympy_mailinglist}        & $\bullet$ &           & $\bullet$ &           &           & $\bullet$ \\
LLVM nuisance-contribution pattern \cite{llvm_policy}                 &           &           & $\bullet$ &           &           & $\bullet$ \\
curl HackerOne bug-bounty shutdown \cite{stenberg_curl_hackerone}     &           & $\bullet$ &           &           & $\bullet$ & $\bullet$ \\
OpenClaw exposed-instance population \cite{strike_openclaw}           &           & $\bullet$ &           &           & $\bullet$ &           \\
\bottomrule
\end{tabular}
\end{table}

Two patterns stand out. \textsc{Maintainer Workload} (D6) is implicated in most cases, from nuisance submissions (LLVM, the SymPy thread) to programme-level shutdowns (curl) to burden on named maintainers (crabby-rathbun); it is the most consequential gap once the unit of analysis is the contribution channel, not the AI system in the abstract. \textsc{Responsibility} (D2) and \textsc{Human Oversight} (D3) stay implicated even when a contribution was rejected: the absence of an accountable party shaped every case, including the post-rejection retaliation that has no current policy or regulatory anchor.

\subsection{Mapping Policies to Regulatory Frameworks}\label{sec:regulatory}

\paragraph{Where open source exceeds regulation.} LLVM's answerability requirement exceeds EU AI Act Article 14, which mandates oversight capacity but not demonstrable understanding, and goes beyond NIST AI RMF \textsc{Govern} by tying accountability to a per-contribution defence rather than a documented role \cite{nistaimrmf}. matplotlib's agent prohibition with named consequences (ban + GitHub report) exceeds the Berkeley \textsc{Govern} function, which specifies the scope of permitted action but no enforcement mechanism \cite{ucberkeley_agentic}. OpenInfra's two-tier labelling operationalises EU AI Act Article 13 and ISO/IEC 42001 Annex A documentation controls at the contribution level, more concretely than either instrument handles system-level transparency \cite{euaiact,iso42001}.

\paragraph{Where regulation provides unmet guidance.} ISO/IEC 23894 requires risk assessment at every lifecycle stage, not only at contributor self-declaration; no policy treats the review, merge, or post-merge stages as risk-assessment loci \cite{iso23894}. ISO/IEC 42001 separates a policy statement (Clause 5.2) from a management system (Clauses 5--10 with Annex A), with documented roles for sponsorship (5.3), risk ownership (6.1), and decommissioning \cite{iso42001}; all six policies sit at the Clause 5.2 level. The Berkeley \textsc{Manage} function and NIST \textsc{Manage}-2 both require containment, including systematic rollback of agent actions \cite{ucberkeley_agentic,nistaimrmf}, yet the crabby-rathbun cleanup was manual across repositories, a \textsc{Manage} failure no policy addresses.

\paragraph{Mutual gaps where neither side addresses.} Three categories of harm are addressed by neither current policies nor regulators. \emph{Maintainer workload as a governance variable}: the EU AI Act, NIST AI RMF, Berkeley Profile, and ISO 42001/23894 all govern providers and deployers of AI systems but ignore what happens to the humans evaluating AI outputs at scale. \emph{Agent-generated harm to third parties who are not users of the system}: EU AI Act Article 5(1)(b) prohibits exploitation of users' vulnerabilities, but maintainers in the matplotlib and SymPy incidents were not users of the crabby-rathbun agent, they were its targets; no policy or regulatory provision yet addresses post-rejection adversarial content production by autonomous agents directed at named individuals. \emph{Anonymous and distributed contribution contexts}: regulatory frameworks assume the deployer is identifiable and legally accountable, while open-source contribution environments frequently involve anonymous or pseudonymous contributors, distributed governance with no single legal entity in control, and platform intermediaries whose ToS create nominal obligations but no reliable downstream enforcement.

\paragraph{A note on the validation cases.} CPython/PSF and SAP, excluded from the primary analysis, sharpen the taxonomy at its boundaries. CPython's policy absence is itself a coding: the observable shape of an unintervened-on policy is \textsc{Absent} on D1 and D2, default-permissive on D3, governed by the project licence on D4, no enforcement mechanism on D5, and zero workload protection on D6. SAP's corporate context shows that the dimensions remain conceptually applicable but the governance instruments differ (employment contracts and CLA frameworks substitute for community norms); the dimensions translate, but per-cell coding categories require domain extension.

%======================================================================
\section{Synthesis: A Proposed Framework}\label{sec:framework}
%======================================================================

The taxonomy and discussion invite a normative successor: a tiered framework that lets small volunteer communities and foundation-stewarded projects share one vocabulary at different operational depths. We sketch its shape rather than a calibrated v1, because the evidence needed to calibrate the tiers does not yet exist. Table~\ref{tab:harmonised} renders three tiers (Minimum Viable, Substantive, Full Alignment) per dimension.

\begin{table}[htbp]
\centering
\caption{Shape of a harmonised three-tier framework. Cells describe the provision suggested at each tier; ordinal calibration is left to future empirical work.}\label{tab:harmonised}
\footnotesize
\setlength{\tabcolsep}{4pt}
\renewcommand{\arraystretch}{1.25}
\newcolumntype{R}{>{\raggedright\arraybackslash}X}
\begin{tabularx}{\textwidth}{@{}>{\raggedright\arraybackslash\bfseries}p{0.19\textwidth} R R R@{}}
\toprule
Dimension & \textbf{Tier 1: Minimum viable} & \textbf{Tier 2: Substantive} & \textbf{Tier 3: Full alignment}\\
\midrule
Disclosure & Recommend disclosure of substantial AI use. & Mandatory disclosure with tool and scope. & Verified two-tier labels (\texttt{Generated-By:}/\texttt{Assisted-By:}) with audit trail.\\
\addlinespace[3pt]
Responsibility & Human responsible for every contribution; agents discouraged. & Explicit autonomous-agent restriction with named consequences. & Documented accountability chain across provider, deployer, contributor, reviewer.\\
\addlinespace[3pt]
Oversight & Human review and understanding expected. & Understanding and demonstrable value-add required. & Answerability test during review without recourse to the AI.\\
\addlinespace[3pt]
Licensing & Flag tool ToU and licence compatibility. & Structured copyright and provenance review. & Code-similarity matching, GPL guidance, audit-grade provenance trail.\\
\addlinespace[3pt]
Enforcement & Maintainer discretion to close low-value AI submissions. & Explicit closure, ban, and platform-report rules. & Auditable enforcement and systematic rollback of agent actions.\\
\addlinespace[3pt]
Workload & Acknowledge reviewer burden; protect onboarding tickets. & Explicit reviewer protection and good-first-issue safeguards. & Rate limits, automated triage, cooldown periods, workload monitoring.\\
\bottomrule
\end{tabularx}
\end{table}

A calibrated v1 should specify recommended ordinal levels per dimension and tier, anchor Tier 3 to simultaneous alignment with the EU AI Act, NIST AI RMF, and ISO 42001, and surface three coordination problems no individual project can solve alone: maintainer workload as a regulated variable (absent from ISO 23894), agent identity verification at platforms (bot-account flagging, signed commits via W3C DIDs, ``autonomous agent'' as a first-class report category), and extension of EU AI Act Article 5(1)(b) to agent-generated content targeting non-users.

%======================================================================
\section{Threats to Validity}\label{sec:threats}
%======================================================================

The analysis is a snapshot of public records through 7 May 2026; later policy iterations may close gaps reported here. Direct evidence for autonomous-agent contribution is concentrated in the crabby-rathbun matplotlib and SymPy cases, and the corpus should not be read as a larger set of autonomous-agent incidents than the public record supports. The incident-to-dimension mapping in Section~\ref{sec:incidents} is a conceptual stress-test correspondence, not a causal claim that a stronger score would have prevented an incident, and the 0--5 rubric is an evaluation instrument that should be replicated with independent coders before being treated as stable. Process tracing depends on public mailing-list threads, PRs, and policy documents; private maintainer discussions and foundation-internal deliberations are not observable. Comparing detailed contributor-facing policies against foundation-level legal guidance requires interpretation, and absences are sometimes structural rather than indicative of weakness. The mapping to the EU AI Act, NIST AI RMF, the Berkeley Agentic Profile, and ISO 42001/23894 is governance analysis rather than legal advice. Finally, SymPy Issue \#29155 remains open and PR \#29156 was closed unmerged as of 7 May 2026: the autonomous-agent gap is best described as stress-tested and unresolved, and any later resolution should be re-coded against the rubric.

%======================================================================
\section{Conclusion and Future Work}\label{sec:conclusion}
%======================================================================

This study presents the first systematic comparative analysis of AI-contribution policies across six major open-source organisations and develops a six-dimensional taxonomy that lets a project locate its governance approach, identify its gaps, and reason about regulatory alignment. Three findings emerge: policy diffusion can strengthen rather than dilute requirements (the disclosure reversal); licensing-focused and oversight-focused policies are distinct archetypes rather than a unified spectrum; and maintainer-workload protection remains universally absent despite empirically documented burden. The regulatory mapping shows these policies sometimes exceed requirements (LLVM's answerability standard) while leaving gaps regulators treat as essential (ISO 23894 lifecycle stages).

Three lines of future work follow. First, a sandbox study deploying autonomous-agent contributions against repositories configured at varying policy strengths, measuring misbehaviour rates, maintainer time, and contribution quality; this is the evidence that would let a v1 framework specify recommended ordinal levels rather than assume them. Second, longitudinal re-coding of the six cases at six-month intervals to test whether the disclosure-reversal pattern generalises. Third, extension to corporate-governance domains (SAP, enterprise CLA frameworks) where the dimensions translate but per-cell categories need domain-specific work. The coding grid is published as supplementary material to support replication.

%======================================================================

\end{document}